\def\be{\begin{equation}}
\def\ee{\end{equation}}
\def\bea{\begin{eqnarray}}
\def\eea{\end{eqnarray}}
\def\bma{\begin{mathletters}}
\def\ema{\end{mathletters}}

\def\0{\overline{0}}

\def\q0{\underline{0}}

\def\one{\leavevmode\hbox{\small1\normalsize\kern-.33em1}}

\documentclass[prl,twocolumn,showpacs,amsmath,amssymb]{revtex4}
\usepackage{dcolumn}                 
\usepackage[dvipdf]{graphicx}       
\usepackage[dvips]{color}
\DeclareGraphicsExtensions{.jpg,.pdf,.mps,.png,.eps,.ps,.EPS}

\begin{document}

\title{The condensation in non-growing complex networks under Boltzmann limit}
\author{Guifeng Su$^{1}$}
\author{Xiaobing Zhang$^{2}$}
\author{Yi Zhang$^{1}$}
\email{yizhang@shnu.edu.cn}
\affiliation{$^{1}$ Department of Physics, Shanghai Normal University, 
Shanghai 200234, People's Republic of China \\
$^{2}$ Department of Physics, Nankai University, Tianjin 300071, 
People's Republic of China}

\begin{abstract}
We extend the Bianconi-Barab\'{a}si (B-B) fitness model to the non-growing 
complex network with fixed number of nodes and links. It is found that the 
statistical physics of this model makes it an appropriate representation of the 
Boltzmann statistics in the context of complex networks. The phase transition
of this extended model is illustrated with numerical simulation and the 
corresponding ``critical temperature'' $T_c$ is identified. We note that the 
``non-condensation phase'' in $T > T_c$ regime is different with ``fit-get-rich'' 
(FGR) phase of B-B model and that the connectivity degree distribution P(k)
deviates from power-law distribution at given temperatures.
\end{abstract}


\maketitle

During the last decade there has been a burst of research activities
on complex networks (see, e.g.,~\cite{Barabasi1, Watts1, Watts2, Mendes1, Pastor1} 
and references therein), from the theoretical modeling as well as the empirical
study of real-world networks. Different approaches and models have already been 
proposed for growing complex networks~\cite{Barabasi2, Arenas1, Donetti1}, and 
the attention has been paid mainly to the out-of-equilibrium dynamics by the 
means of preferential attachment~\cite{Barabasi3, Newman1, Barabasi4, 
Dorogovtsev1, Bianconi1}. Of particular interest is the Bianconi-Barab\'{a}si 
(B-B) fitness model~\cite{Bianconi1} and the corresponding condensate behavior 
in the growing networks~\cite{Bianconi2}.

In the fitness model, each nodes of the network corresponds to different energy 
levels of a system through the relation
\begin{equation}\label{energylevel}
    \epsilon_i=-\frac{1}{\beta}\log \eta_i ,
\end{equation}
where $\eta_i$ is the fitness of node $i$, and $\beta$ is an inverse
temperature parameter (i.e., $\beta = 1/T$). The probability that a
new link connects to the $i$th node depends on the fitness (energy
level) $\eta_i$ and degree of connectivity (the number of links) of the $i$th 
node $k_i$,
\begin{equation}\label{probability}
    \Pi_i=\frac{k_i e^{-\beta\epsilon_i}}{\sum_jk_j e^{-\beta\epsilon_j}} .
\end{equation}

In Ref.~\cite{Bianconi2}, the evolving network with the fitness model was mapped 
into an equilibrium Bose gas. With mean-field arguments, the well-known 
Bose-Einstein (BE) statistics of the occupation number of an energy level (with
energy $\epsilon)$, $n(\epsilon) = 1/(e^{\beta (\epsilon - \mu)} - 1)$, was 
obtained in the thermodynamic limit ($t \rightarrow \infty$). As a result, the 
network may undergo distinct phases -- so called ``scale-free'' phase, 
``fit-get-rich'' (FGR) phase and Bose-Einstein-condensate (BEC) phase. The 
appearance of BEC phase indicates a ``winner-takes-all'' phenomenon -- the fittest 
node has a finite fraction of the total number of links -- during the evolution 
of the network.

Although mapping to Bose gases successfully, one notices that the non-equilibrium
(e.g., the growth of the number of nodes and the inertness of the links), and 
irreversible natures make B-B model an unreal representation of BE statistics of 
Boson quantum gases. This motivates us to extend the fitness model to a non-growing 
network with the fixed number of nodes and links.

This non-growing network with fitness is constructed in the following way: initially, 
a graph composed of $N$ nodes and $K$ links ($N>K$) is given, and each node is 
assigned an energy $\epsilon_i$ chosen from some energy level distribution $g(\epsilon)$. 
We then let the network evolve: at each time step, a link is randomly picked up and
disconnected one end while the other end is kept fixed, then this link is rewired to 
a new node with the probability $\Pi_i$ (see eqn.~(\ref{probability})) until time steps 
reach the number of links $K$. The corresponding rate equation for the degree of
connectivity of the $i$th node, $k_i$, read as
\begin{equation}\label{rate}
    \frac{\partial{k_i}}{\partial{t}}=\frac{k_i(\epsilon_i,t,k_{i0}) e^{-\beta\epsilon_i}}{Z_t}-\frac{k_i(\epsilon_i,t,k_{i0})}{K}
    , i=1,2,\cdots,N
\end{equation}
where
\begin{equation}
    Z_t=\sum^{N}_{j=1}k_j(\epsilon_j,t,k_{j0})e^{-\beta\epsilon_j} ,
\end{equation}
is the partition function, and $k_{i0}$ is the initial distribution of 
connectivity degree at time $t=0$. Eqn.(\ref{rate}) is a group of $N$ coupled 
differential equations, in principle, we can solve it and determine the degree 
of connectivity $k_i$. However, a qualitative analysis is of interest here: 
first, the whole procedure can be seen as a mapping to Bose gases but with the 
characteristics of equilibrium evolution and the rewiring ability of links (so 
avoid the inertness in B-B model); second, in the low temperature regime, the 
rate equation can be simplified to
\begin{equation}\label{ratepr}
    \frac{\partial{k_1}}{\partial{t}}\approx\frac{k_1(\epsilon_0,t,k_{10}) e^{-\beta\epsilon_1}}{Z_t} ,
\end{equation}
\begin{equation}\label{ratenonz}
    \frac{\partial{k_j}}{\partial{t}}<0 , j\geqslant2 .
\end{equation}
where $k_1$ is the degree distribution of the lowest energy level.

The form of eqn.~(\ref{ratepr}) is similar to eqn.~(3) in Ref.~\cite{Bianconi2}. 
However, it is noteworthy that there, eqn.~(3) holds for all energy levels 
(and henceforth rusults in a BE statistics). Following eqn.~(\ref{ratepr}) 
and eqn.~(\ref{ratenonz}), the links are eventually 
disconnected from higher energy levels, and rewired to the lowest energy level, 
this corresponds to a condensation phenomenon~\cite{Remark1}, and $k_1$ saturates 
when the two terms of right hand side (rhs.) of eqn.~(\ref{rate}) cancels.


The scenario of this non-growing network under the classical limit is 
very interesting and deserves detailed investigations. Under the classical 
limit, the links (or particles) are independent of each other, and the 
connectivity probability of links depend only on the fitness $\eta_i$ 
or energy level $\epsilon_i$, but not on the degree of connection of the
$i$th node, $k_i$. We want to emphasize that this difference in the 
probability of connection makes the whole phase structure very different 
with FGR phase in the growing networks, as what we are going to explain 
later with our simulations. In the classical limit the connectivity 
probability of network becomes
\begin{equation}\label{probpr}
    \Pi_i=\frac{e^{-\beta\epsilon_i}}{\sum_j e^{-\beta\epsilon_j}}=\frac{e^{-\beta\epsilon_i}}{Z} ,
\end{equation}
where
\begin{equation}\label{partifcl}
 Z=\sum_i e^{-\beta\epsilon_i} .
\end{equation}
In classical limit, this non-growing network constructed in a given 
``temperature'' $\beta$ corresponds to an equilibrium gas composed of 
$K$ independent particles distributed in $N$ energy levels. According 
to the standard statistical physics, for chemical potential $\mu$ and 
the partition function $Z$ \cite{KHuang}, we have
\begin{equation}
\label{muandz}
\mu = -\frac{1}{\beta}\ln\frac{Z}{K}.
\end{equation}
combine (\ref{partifcl}) and (\ref{muandz}) we get
\begin{equation}\label{10}
Z = \sum_i e^{-\beta \epsilon_i} = K e^{-\beta \mu} = \sum_i k_i e^{-\beta \mu}.
\end{equation}
Eqn.(\ref{10}) simply leads to
\begin{equation}
K = \sum_i k_i=\sum_i e^{-\beta(\epsilon_i-\mu)},
\end{equation}
which is, clearly, Boltzmann statistics with distribution $f(\epsilon_i)$, i.e.,
\begin{equation}
k_i \sim f(\epsilon_i) = e^{-\beta(\epsilon_i-\mu)}.
\end{equation}

In order to show the phase transition of this non-growing network, we 
numerically simulate the above scenario under the classical limit. we 
generate a graph consist of $N$ nodes and $K$ links. These links are 
rewired to existing nodes by the following way: one point of every 
link connect to one node $j$ randomly while the other point connect to 
another node $i$ with probability $\Pi_i$ by eqn.~(\ref{probpr}). The 
situation that two or more links connecting the same pair of nodes is 
not allowed.  

\begin{figure}
\begin{center}
  \includegraphics[width=8cm]{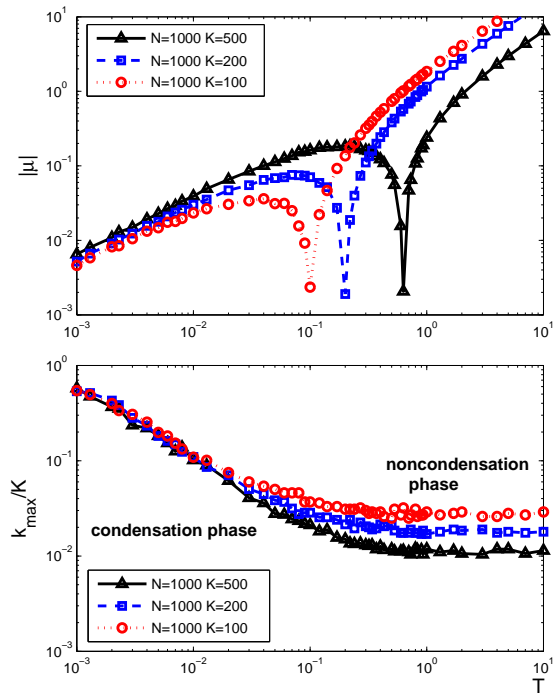}\\
  \caption{Simulation results of the network under Boltzmann limit for 
  distribution $g(\epsilon) = C \epsilon^\theta$ with $\theta = 0$. The 
  upper panel is the plot of absolute value of chemical potential $\mu$ 
  with temperature $T$ on a logarithmic scale, the lower panel is the 
  fraction of total links occupied by the most connected nodes. The 
  curves are plotted for the same number of nodes $N = 1000$, but 
  different number of total links: $K = 500$ (open triangle with solid 
  line), $K = 200$ (open square with dashed line), and $K = 100$ (open 
  circle with dotted line), respectively. The numerical simulations are 
  averaged over 50 runs.}
\label{fig1}
\end{center}
\end{figure}

\begin{figure}
\begin{center}
  \includegraphics[width=8cm]{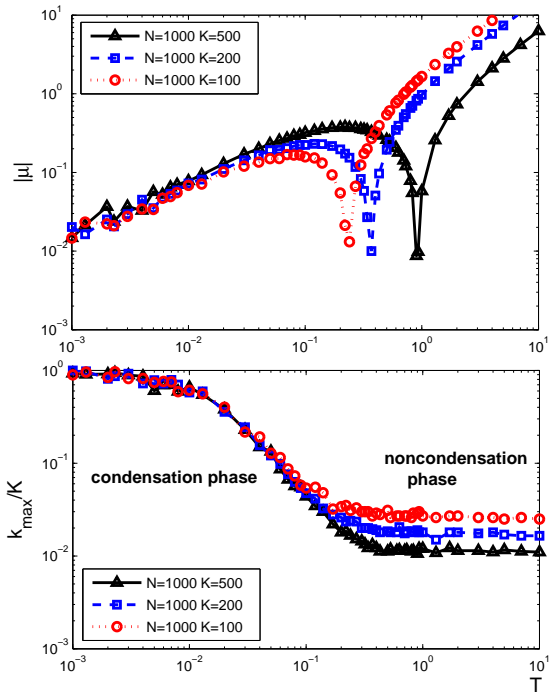}\\
  \caption{Similar to the Fig.~\ref{fig1}, but for the distribution of energy level 
   $g(\epsilon) = C\epsilon^\theta$ with $\theta = 1$. }
\label{fig2}
\end{center}
\end{figure}

\begin{figure}
\begin{center}
  \includegraphics[width=8cm]{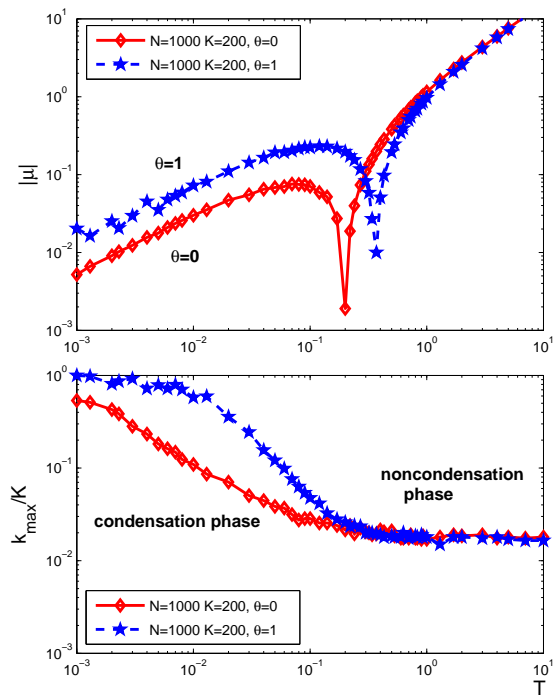}\\
  \caption{The comparisons of the absolute value of chemical potential, $|\mu|$ 
  (upper panel), and $k_{max}/K$ (lower panel) with the same number of nodes 
  ($N = 1000$) and links ($K = 200$), but with different energy level distributions
  $g(\epsilon) = C \epsilon^\theta$, $\theta = 0$ (open diamond with solid lines), 
  and $\theta=1$ (star with dashed lines).}
\label{fig3}
\end{center}
\end{figure}

\begin{figure}
\begin{center}
  \includegraphics[width=8cm]{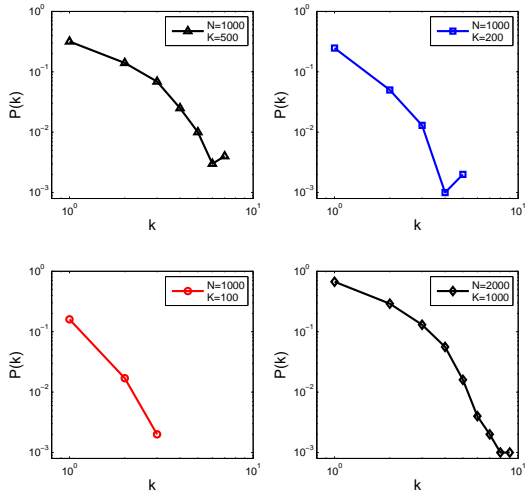}\\
  \caption{The connectivity degree distribution $P(k)$ vs. degree $k$ for 
  $N = 1000$, $K = 500$ (open triangles), $200$ (open squares), $100$ 
  (open circles) and $N = 2000$, $K = 1000$ (open diamonds) with distribution 
  $g(\epsilon) = C \epsilon^\theta$, $\theta = 0$ at temperature $T = 0.2$.}
\label{fig4}
\end{center}
\end{figure}

\begin{figure}
\begin{center}
  \includegraphics[width=8cm]{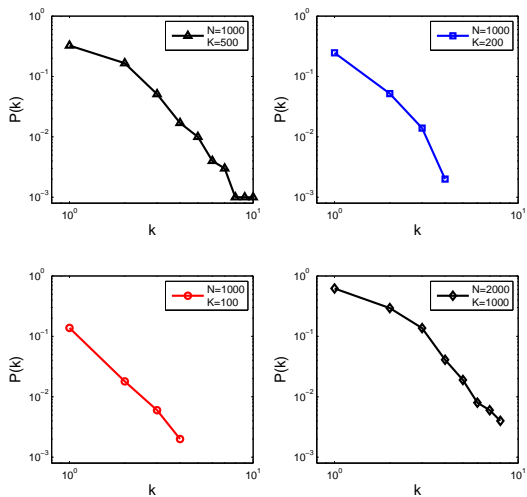}\\
  \caption{Similar to the Fig.~\ref{fig4} but for the distribution of energy levels 
  $g(\epsilon) = C \epsilon^\theta$ with $\theta = 1$.}
\label{fig5}
\end{center}
\end{figure}

The simulation results of our model in the classical limit are shown in 
Figs.~\ref{fig1} -- \ref{fig5}. In Fig.~\ref{fig1}, we show the 
relationship of absolute value of chemical potential $|\mu|$ with 
temperature parameter $T=1/\beta$, in upper panel. The curves represent 
networks with $N = 1000$ and $K = 500$, $200$, $100$, respectively. It 
can be seen clearly that there exist transitions in $|\mu|$, which means 
the chemical potential $\mu$ changes its sign, i.e., $\mu$ is positive 
in $T < T_c$ regime and negative in $T > T_c$ regime. The negativeness 
of $\mu$ in $T > T_c$ regime only has a formal similarity to that of the 
ideal Boltzmann gases, since, after all, the classical limit condition 
is not satisfied in the low temperature regime and hence no such 
condensation behavior in Boltzmann statistics.

The (non)condensation behavior is shown in the lower panel in 
Fig.~\ref{fig1}. We can see that the fraction increases to unity gradually, 
as temperature decreases. Also, the critical temperature $T_c$ shifts 
towards to higher temperature when the ratio $K/N$ increases. In low 
temperature limit, the connectivity probability of non-growing network 
is primarily dominated by factor $e^{-\beta \epsilon_i}$ rather than $k_i$, 
which is similar to a network under the classical limit. As we already 
mentioned earlier, it should be noted that \emph{the non-condensation 
phase of classical limit network is different from the so called ``FGR'' 
phase of the fitness model in growing network. In non-condensation phase 
the connectivity probability of each node is randomly distributed over all 
energy levels. However, in FGR phase, for the nodes with higher fitness 
the connectivity degree increase more quickly than those nodes with lower 
fitness.}

Different distributions of energy level, $g(\epsilon)$, also influence the 
(non)condensation phase behavior under the classical limit. In our model, we 
employ the following form,
\begin{equation}\label{gepsilon}
    g(\epsilon) = C\epsilon^\theta ,
\end{equation}
where $\theta$ is a parameter and $C$ is normalization factor. The simulations 
shown in Fig.~\ref{fig1} correspond to $\theta = 0$, $C = 1$, {\it i.e.}, a 
uniform distribution. In Fig.~\ref{fig2}, we show the simulation results of 
$|\mu|$ and $k_{max}/K$ for $\theta = 1$, $C = 2$. To illustrate clearly the 
effects of $g(\epsilon)$, in Fig.~\ref{fig3} 
we show the comparisions of the simulations with $\theta = 0$ and $\theta = 1$ for
the same number of nodes and links ($N = 1000$, $K = 200$). The enhancement of critical 
transition temperature $T_c$ for distribution with $\theta = 1$ is obvious.

We also plot the probability distribution of connectivity degree $P(k)$ at 
temperature $T = 0.2$ for distributions with $\theta = 0$ (Fig.~\ref{fig4}) and 
$\theta = 1$ (Fig.~\ref{fig5}), respectively. The up left, up right and lower left 
panels in Fig.~\ref{fig4} and Fig.~\ref{fig5} correspond to the following number of nodes 
and links: $N = 1000$ and $K = 500$, $200$, $100$. With different number 
of nodes $N$ and links $K$ but the same ratios of $K/N$ as shown in the up 
left and lower right panels in Fig.~\ref{fig4} and Fig.~\ref{fig5}, respectively, the 
distributions $P(k)$ (vs. $k$) are quite similar except for the fluctuation 
in tails. A particularly interesting observation is, at given temperature, 
it seems like the probability distributions of $k_i$ deviate from power-law 
distribution, as with the increase of the ratios $K/N$, which is clearly 
demonstrated in both figures. This significant difference from scale free 
network show some hints on the possible connection with recent observations: 
some real-world complex networks display a Weibull- or mixed-Weibull-power-law
distribution~\cite{Xu, He07, Chen, Chang, Leskovec, Seshadri}. Although some
models are proposed to explain (see, e.g., Ref.~\cite{He07}), the underlying 
mechanism for such distribution is worthy of further studying.  

In conclusion, in present paper, we extend Bianconi-Barab\'{a}si fitness 
model from growing complex networks to non-growing networks with fixed number 
of nodes and links. We fulfill numerical simulations and identify the 
corresponding ``critical temperature'' $T_c$. We also find that the 
``non-condensation phase'' of our extension differs from FGR phase in B-B 
model, and the degree distribution $P(k)$ deviates from the power-law 
distribution at some temperatures.

\noindent{\bf Acknowledgements.}
The authors thank Prof. Albert-L. Barab\'{a}si for discussions and acknowledge 
to National Natural Science Foundation of China (NSFC) for their support, under 
Contract No. 10875058.



\begin{thebibliography}{999}

\bibitem{Watts1} D. J. Watts and S. H. Strogatz, Nature (London) 393, 440 (1998).

\bibitem{Barabasi1} R. Albert and A.-L. Barab\'{a}si, Rev. Mod. Phys. 74, 47
(2002).

\bibitem{Watts2} D. J. Watts, {\it Small Worlds: The Dynamics of Networks between Order 
and Randomness}, Princeton University Press, Princeton, NJ, 1999.

\bibitem{Mendes1} S. N. Dorogovtsev and J. F. F. Mendes, {\it Evolution of Networks}, 
Oxford University Press, Oxford, 2003.

\bibitem{Pastor1} R. Pastor-Satorras, A. Vespignani, {\it Evolution and Structure 
of the Internet: A Statistical Physics Approach}, Cambridge University Press,
Cambridge, 2004.

\bibitem{Barabasi2} R. Albert and A.-L. Barab\'{a}si, Phys. Rev. Lett. 85, 5234 (2000).

\bibitem{Arenas1} R. Guimer\'{a}, A. D\'{\i}az-Guilera, F. Vega-Redondo, A.
Cabrales and A. Arenas, Phys. Rev. Lett. 89, 248701 (2002).

\bibitem{Donetti1} L. Donetti, P. I. Hurtado and M. A. Mu\~noz, Phys. Rev. Lett. 95, 
188701 (2005). 

\bibitem{Barabasi3} A.-L. Barab\'{a}si and R. Albert, Science 286, 509 (1999).

\bibitem{Newman1} M. E. J. Newman, Phys. Rev. E 64, 025102 (2001).

\bibitem{Barabasi4} H. Jeong, Z. Neda, A.-L. Barab\'{a}si, Europhys. Lett. 61, 
567 (2003).

\bibitem{Dorogovtsev1}  S. N. Dorogovtsev, J. F. F. Mendes and J. G. Oliveira, 
Phys. Rev. E 73, 056122 (2006).

\bibitem{Bianconi1} G. Bianconi and A.-L. Barab\'{a}si, Europhys. Lett. 54, 436-442 (2001).

\bibitem{Bianconi2} G. Bianconi and A.-L. Barab\'{a}si, Phys. Rev. Lett. 86, 5632
(2001).

\bibitem{Remark1} At least for the lowest energy level, there exists the
possibility of a BEC phase transition. However, whether or not BE statistics holds 
for \emph{all energy levels}, is still under investigations~\cite{SZZ10}.

\bibitem{KHuang} K. Huang, Statistical Mechanics, Wiley, Singapore, 1987.

\bibitem{Xu} K. Xu, L. Liu and X. Liang, arXiv:0908.0588 [cond-mat].

\bibitem{He07} Y. He, G. Siganos, M. Faloutsos, S. V. Krishnamurthy, 
Proc. USENIX/SIGCOMM NSDI (2007).

\bibitem{Chen} Q. Chen, H. Chang, R. Govindan, S. Jamin, S. Shenker 
and W. Willinger, Proc. INFOCOM (2002).

\bibitem{Chang} H. Chang, S. Jamin and W. Willinger, Proc. INFOCOM (2006).

\bibitem{Leskovec} J. Leskovec and E. Horvitz, Proc. WWW (2008).


\bibitem{Seshadri} M. Seshadri, S. Machiraju, A. Sridharan, J. Bolot, C. Faloutsos 
and J. Leskove, Proc. SIGKDD (2008).

\bibitem{SZZ10} G. Su, Y. Zhang and X. Zhang, In preparation.




































\end{thebibliography}
\end{document}